# Coexistence of charged stripes and superconductivity in $Bi_2Sr_2CaCu_2O_{8+\delta}$

C. Howald,* H. Eisaki,† N. Kaneko,‡ & A. Kapitulnik*†

** Department of Physics, Stanford University, Stanford, CA 94305, USA*

† *Department of Applied Physics, Stanford University, Stanford, CA 94305, USA*

‡ *Stanford Linear Accelerator Center, Menlo Park, CA 94025, USA*

**One of the most important outstanding problems in condensed matter physics is determination the ground state of strongly correlated electron system, in particular high-temperature superconductors (HTSC). For these materials, theoretical[1-6] and experimental[7-13] evidence has been mounting in support of the possibility that their ground state exhibits spin and charge density waves (SDW and CDW), which are primarily one-dimensional (i.e. stripes) and which either compete with or promote superconductivity. Coexistence of CDW or SDW and superconductivity has previously been reported in the lower $T_c$ materials[7-9,12,13] and in the presence of large magnetic fields[10,11]. The absence of data showing stripes in the higher $T_c$ materials in zero field has lent support to the idea that stripes are competing with high temperature superconductivity. However, here we show, using scanning tunneling spectroscopy (STS), the existence of static striped density of electronic states in nearly optimally doped $Bi_2Sr_2CaCu_2O_{8+\delta}$ in zero field. This modulation is aligned with the Cu-O bonds, with a periodicity of four lattice constants, and exhibits features characteristic of a two-dimensional system of line objects. We further show that the density of states modulation manifests itself as a**

**shift of states from above to below the superconducting gap. The fact that a single energy scale (i. e., the gap) appears for both superconductivity and stripes suggests that these two effects have the same origin.**

STS allows one to measure, on an atomic scale, the electronic density of states. This measurement of the local density of states (LDOS) makes it a powerful tool for investigating correlated electron systems. In this study we used $Bi_2Sr_2CaCu_2O_{8+\delta}$ because it cleaves easily, yielding large, stable, atomically flat surfaces and has a high $T_c$(~ 90 K). The single crystals of $Bi_2Sr_2CaCu_2O_{8+\delta}$ used in this study, grown by a floating-zone method, were annealed to be slightly overdoped, yielding a superconducting transition temperature of 86 K. Fig. 1a shows the topography of a typical area: the surface exhibits clear atomic resolution. In particular, the superstructure in the BiO plane[14] with average periodicity ~27 Å (as well as the location of the individual Bi atoms) is evident in all scans and provides a reference direction for our study. As reported earlier[15-17], all samples also exhibit strong inhomogeneity in the local spectroscopy. This manifests itself here as large variation in the superconducting gap size, as measured by the location of the peak in the local density of states (Fig. 1b). The largest gaps ($\Delta$ > 60 mV) account for ~10% of the area. The variation in the spectra is shown in the line scan of Fig. 1c, exhibiting evolution from small superconducting gap ($\Delta$ < 40 mV) with large coherence peaks to regions of large gap with much reduced coherence peaks reminiscent of the pseudogap[18,19].

To facilitate our search for charge modulations in the superconducting state, we collected LDOS spectra at each point in the image, and subsequently generated LDOS maps at various energies and their corresponding power spectra (intensity of the Fourier transform); one such pair is shown in Figs. 2a and 2b for sample bias of 15 mV. There is a large amount of weight at small wavevectors, coming from the random contributions of the large inhomogeneities. In addition, there are two bright peaks appearing along the $k_x$ direction which originate from the superstructure. Finally, Fig. 2a clearly shows four peaks, oriented 45 degrees to the superstructure and at points corresponding to, within experimental uncertainty, periodicities of $(\pm 4a_0, 0)$ and $(0, \pm 4a_0)$, oriented along the copper-oxygen bonds. The real-space data (Fig. 2b) show a two-dimensional pattern in the LDOS that is oriented along the diagonals; however, as is evident from Fig. 2a, contributions from features at other wavelengths obscure this modulation. We can use Fig. 2a to Fourier filter out from Fig. 2b contributions to the LDOS that are far away from the $(2\pi/4a_0)(\pm 1,0)$ and $(2\pi/4a_0)(0,\pm 1)$ points. Fig. 2c shows the data after such Fourier filtering with a standard gaussian filter function using $\sigma = (2\pi/15a_0)$. Varying the width of this filter has no qualitative effect on the resulting image. The modulation partially visible in Fig. 2b is now clear, appearing as a somewhat disordered 'checkerboard' pattern. From the separation of the defects in this map we estimate the coherence length to be $25a_0$.

Recent experiments on several HTSC systems have found similar charge or spin modulations in the presence of high magnetic fields applied perpendicular to the $CuO_2$ layers. From general theoretical arguments, it is expected that for coupled spin and charge density waves, spin modulation will have twice the period of charge modulation[1-6]. In inelastic neutron scattering experiments on $La_{2-x}Sr_xCuO_4$ near optimum doping (x=0.163), Lake et al.[8-9] found strong scattering peaks at the four k-space points: $(\pi/a_0)[(1/2,1/2) \pm \delta(0,1/2)]$ and $(\pi/a_0)[(1/2,1/2) \pm \delta(1/2,0)]$, where $\delta \sim 0.25$ and $a_0$ is the lattice constant. This implies a spin-density of periodicity $\sim 8a_0$, which the authors find extends to more than $50a_0$, much beyond the vortex core.  Also, Mitrovic et al.[10] reported similar periodicity in high-field NMR imaging experiment. In a remarkable experiment, Hoffman et al.[11] reported scanning tunneling spectroscopy on $Bi_2Sr_2CaCu_2O_{8+\delta}$ single crystals revealing a 'checkerboard' of quasiparticle states with four unit-cell periodicity surrounding vortex cores. In k-space this corresponds to Bragg peaks in the local density of states at $(2\pi/a_0)(\pm 1/4,0)$, and $(2\pi/a_0)(0,\pm 1/4)$. This structure was found around the energy (~7 meV) of a feature previously observed in the spectra of vortices in the same field range[20,21]. The 'checkerboard' structure extends over ~ $20a_0$, again indicating ordering outside the vortex cores[22].

We have also found that the modulation in the LDOS exhibits strong energy dependence (Fig. 3). Here we show the value of the Fourier transform at the two independent peaks from Fig. 2a as a function of the bias voltage. The arbitrary phase (dependent on the point about which the Fourier transform is computed) has been set to maximize the real part, plotted in red, which changes sign at about ± 40 mV. The variation in the imaginary part of the signal is considerably smaller, consistent with zero within the uncertainty. This is important since it shows that the location in real-space of the density of states (DOS) modulation is the same at all measured energies, although at about ± 40 mV the positions of the maxima and the minima switch.

Furthermore, the energy dependence of the Fourier transform shows that relative spectral weight is shifted to sub-gap energies (with a peak at ±25 mV) from intermediate energies (with weight between 50 and 150 mV). The comparison of the above result with the energy of the superconducting gap ($\Delta$ ~ 40 mV) provides strong circumstantial evidence that the striped quasiparticle density and superconductivity are intimately connected. Since the integrated density of states from zero to infinity is the total charge, the shift of weight from intermediate to low energies coupled with the small magnitude of the Fourier components (< 5% of the LDOS) indicates that the total charge associated with this density of states modulation is very small, as is expected for itinerant systems with strong Coulomb interactions[23-25]. On the other hand, it is important to note that energy independent spatial variations in the DOS would not appear in this analysis because the normalization is set by maintaining constant current (which is proportional to the integrated DOS from zero to the setpoint voltage). We are sensitive to features in the shape of the LDOS, not the overall magnitude. Energy independent changes would however appear in the so-called topographic signal (Fig. 1a), which actually maps a contour of constant integrated density of states. However, the power spectrum of the topograph reveals no such peaks above the noise.

To examine whether the LDOS modulation is anisotropic, we performed Fourier analysis on many smaller sub-regions of a slightly larger area. In Fig. 4c and 4d we show maps of the spatial variation of the amplitude of the two perpendicular pairs of Bragg peaks. In addition to the small scale variations which come from noise and finite size effects, it is clear that each varies on the scale of roughly 100 Å. In addition, these large scale features exhibit significant anticorrelations between the two maps: there are several regions in space where one map shows a local maximum while the other shows a local minimum. This indicates that locally there is a broken symmetry, and that the underlying order is not two-dimensional, but one-dimensional. The two-dimensional nature of the patterns observed appears to come from the interpenetration of these perpendicularly oriented stripes.

Following the work of Zaanen and Gunnarson[1] and of Schulz[26] which first suggested spatial segregation in HTSC, Kivelson and Emery[2] pointed out that disordered or fluctuating metallic stripe phases are a natural occurrence in doped Mott-Hubbard insulators. In this scenario, charge-ordered phases will then be formed. This charge ordering in turn drives the modulation of the antiferromagnetic order, hence resulting in an accompanying spin density wave structure. Near optimal doping, it is argued, a mixed state exists with both charge and spin density waves. While in general stripes are expected to be dynamic in this regime, disorder will easily pin them[2,4,27]. Vortex-induced pinning of stripes was used by Hoffman et al. [9] to explain the observed density modulation near vortex cores. Random point defects, present in all HTSC samples, could play a similar role, thus explaining the origin of the stripe patterns we observe.

In addition to this static striped density of states structure that spans the entire surface, gap structure inhomogeneities are observed in all our samples. The piling of LDOS in mid-gap, around a particular energy, and their origin from states just above the gap suggest that the density modulation and superconductivity are inseparable. The inhomogeneities in the superconducting gap structure should therefore also be reflected in the striped density of states. Indeed, the striping does show local variation as is shown in Fig. 4, together with topological defects in the striped structure visible in Fig. 2c[28]. Since in our samples the disorder seems to be uncorrelated, it cannot establish a preferred direction, something else must be breaking rotational symmetry. We further note that since the observed stripes are static, yet superconductivity is strong, these stripes cannot be inherently insulating. Finally, it is remarkable that in this single set of experiments we see locally the d-wave superconducting gap, the pseudogap, and metallic stripes, suggesting that all reflect aspects of the same physics.

Acknowledgements: We thank Steven Kivelson and Subir Sachdev for many helpful discussions. We particularly want to thank Steven Kivelson for critical reading of the manuscript. We are indebted to Prof. Martin Greven for the use of the growth system. This work was funded by the AFOSR. HE acknowledges support by the Chodorow Fellowship, Stanford University. NK acknowledges support by the DoE.

**Figure 1** Inhomogeneity in the electronic structure of a slightly overdoped $Bi_2Sr_2CaCu_2O_{8+\delta}$ single crystal.

**a,** Topography (160 Å x 160 Å) of the cleaved BiO surface (spatial units are in angstroms throughout). The samples were cleaved (between the BiO planes) at room temperature in a vacuum of better than $10^{-9}$ torr, then transferred in less than one minute to the low temperature STM, where cryopumping yields orders of magnitude better vacuum. Data were taken at 8 K with an iridium tip at a sample bias of -200 mV and a setpoint current of -100 pA. These setpoints also establish the relatively arbitrary normalization condition for the differential conductance (dI/dV), proportional to the local density of states. The height is a contour of constant current, or integrated density of states up to the setpoint voltage. The nearly vertical streaks are the superstructural modulation. Also visible are the Bi atoms, as well as irregular variations that are probably due to changes in the LDOS[16], not actual height variation. **b,** The superconducting gap magnitude (Δ) over the same area, as measured by the voltage of the maximum in differential conductance (dI/dV). **c,** Differential conductance as a function of voltage along the diagonal of **a** and **b**, from lower left to upper right. Each spectrum is colored according to its measured gap magnitude, using the same color scale as **b**.

**Figure 2** Periodic spatial variation in the quasiparticle density of states.

**a,** Power spectrum of the Fourier transform of dI/dV(15 mV). The units are inverse angstroms and correspond to inverse wavelength. The inset shows a schematic of the features in the power spectrum; the arrows show the Cu-O-Cu directions. The central red dot corresponds to the signal at small wavevectors comes from the randomly distributed inhomogeneities shown in Fig 1b,c. The horizontal points, which are green in the inset, correspond to the superstructure. The four diagonal points, yellow in the inset, show a density of states modulation at **k**=$(2\pi/4a_0)(0, \pm1)$ and $(2\pi/4a_0)(\pm1,0)$. **b,** Map of dI/dV(+15 mV), smoothed by averaging neighboring pixels to remove atomic scale variations. **c,** The data in **b**, filtered to accentuate the periodic modulation. The filter is performed by multiplying the Fourier transform of the raw data by a sum of gaussian weighting functions with width $\sigma=(2\pi/15a_0)$ centered about the four diagonal points in **a**, then applying the inverse Fourier transform.

**Figure 3** Energy dependence of the periodic local density of states modulation.

Plot of the Fourier transform at k = $(2\pi/4a_0)(0, \pm 1)$ and $(2\pi/4a_0)(\pm 1,0)$, the location of the peaks in Fig. 2**a**, as a function of sample bias. The red and blue traces correspond to the real and imaginary parts, respectively. For each peak, the overall phase used for all energies is chosen to maximize the real part of the signal. The error bars are determined by modeling the variations in dI/dV(x,y) at each energy by uncorrelated noise of the same amplitude.

**Figure 4** Spatial variation in the amplitude of the periodic modulation.

**a,** Constant current topograph (130 Å x 130 Å). **b,** Power spectrum of dI/dV(15 mV) over the entire 260 Å x 260 Å area showing the peaks used in **c** and **d**. Peaks are smaller than those of Fig. 2a because the k-space pixels are smaller and because the image size substantially exceeds the coherence length. **c,** Magnitude of the Fourier transform of dI/dV(15 mV) at one of the $(2\pi/4a_0)$ points for a number of 32 x 32 pixel regions. The color at each point is the magnitude of this Fourier component for the 32 x 32 pixel dI/dV map centered at this point. The small (~ 20 Å) scale variations are due to finite size effects in the Fourier transform. **d,** The same map as **c**, for the other peak.

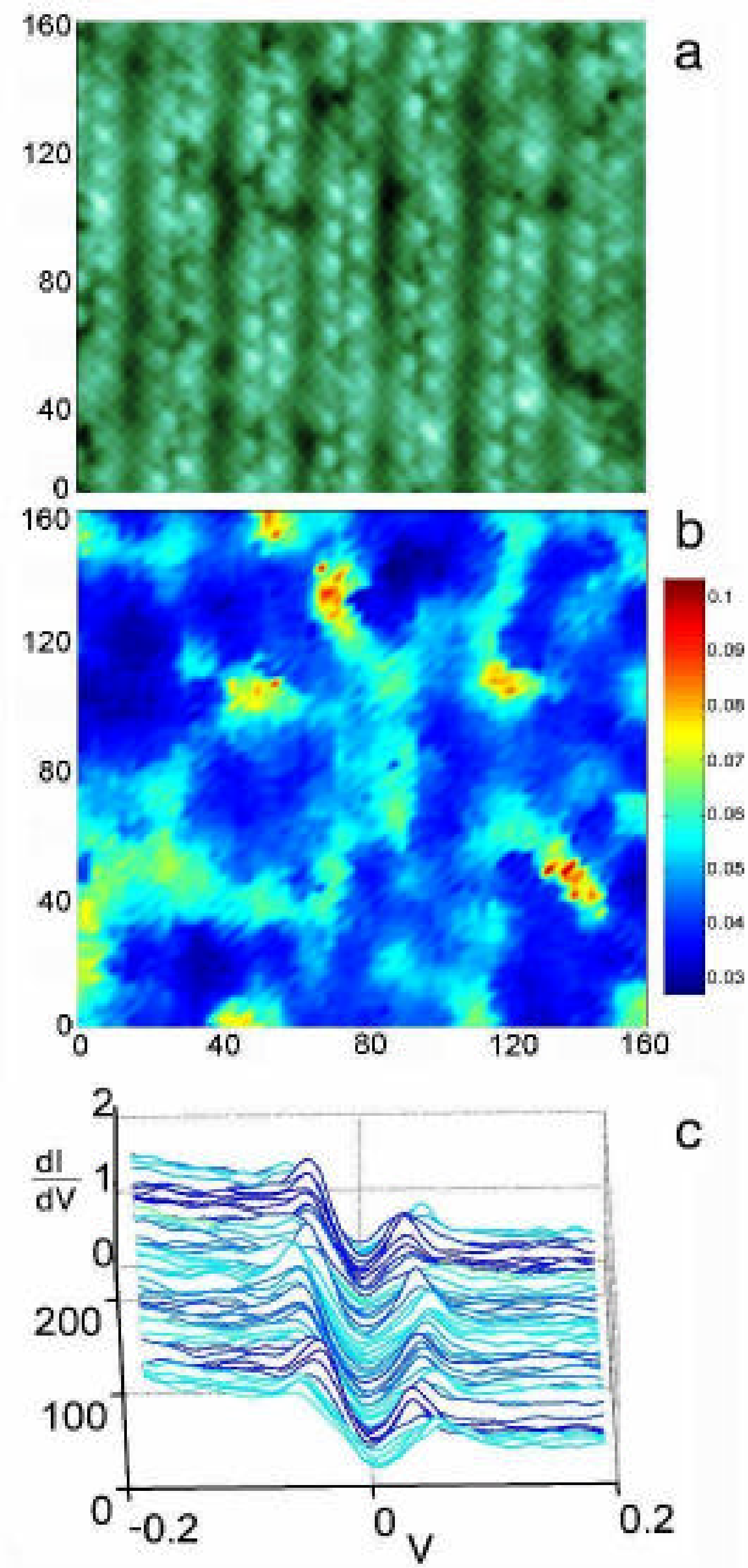
a
b
c
160
120
80
40
0
0.1
0.09
0.08
0.07
0.06
0.05
0.04
0.03
0
40
80
120
160
2
1
0
dI
dV
200
100
0
-0.2
0
0.2
V

Figure 1

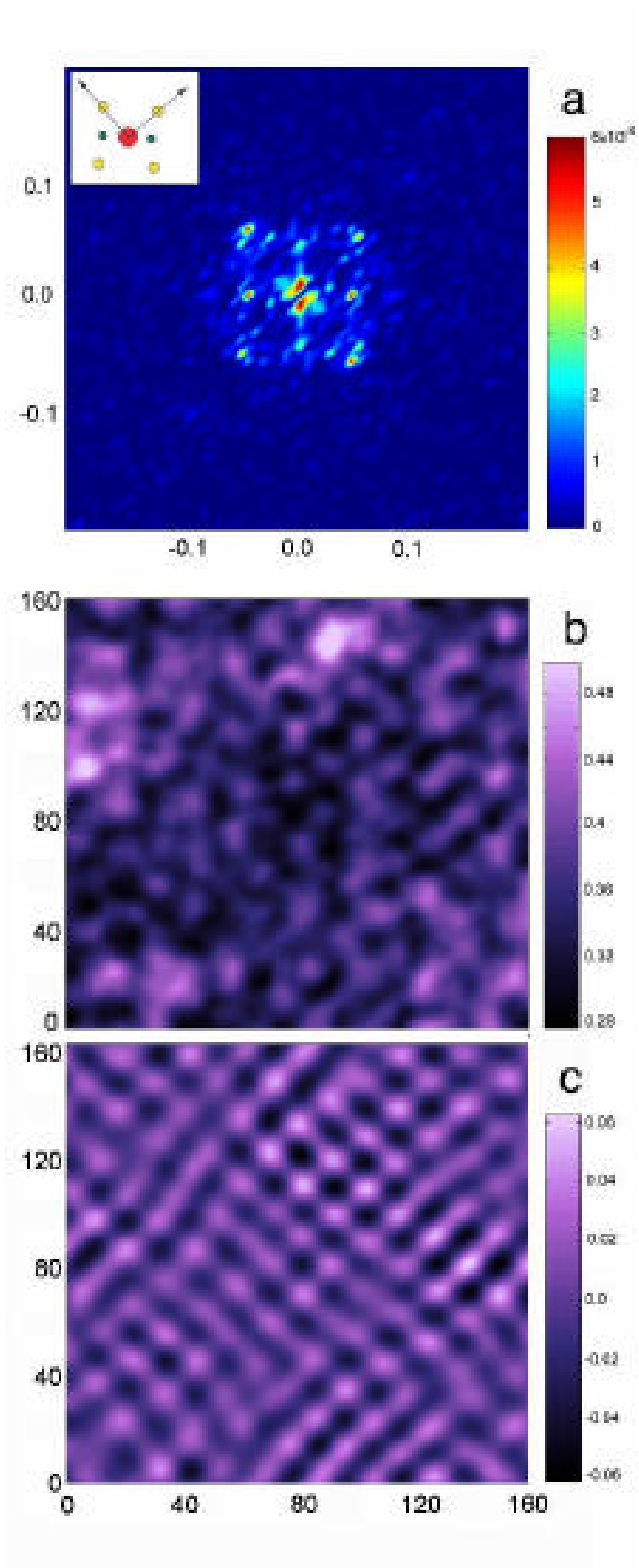

Figure 2

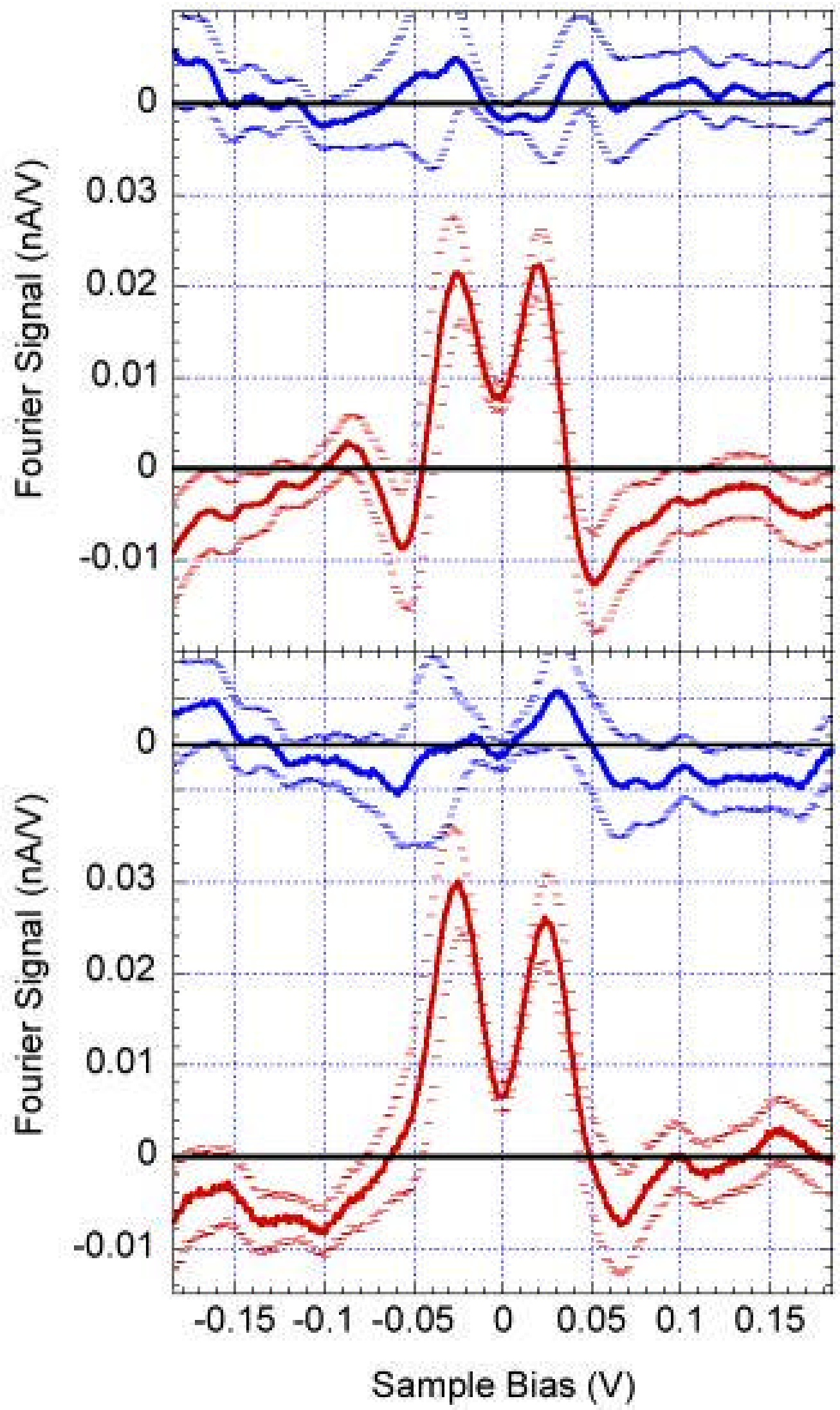

0
0.03
0.02
0.01
0
-0.01
Fourier Signal (nA/V)
0
0.03
0.02
0.01
0
-0.01
Fourier Signal (nA/V)
-0.15 -0.1 -0.05 0 0.05 0.1 0.15
Sample Bias (V)

Figure 3

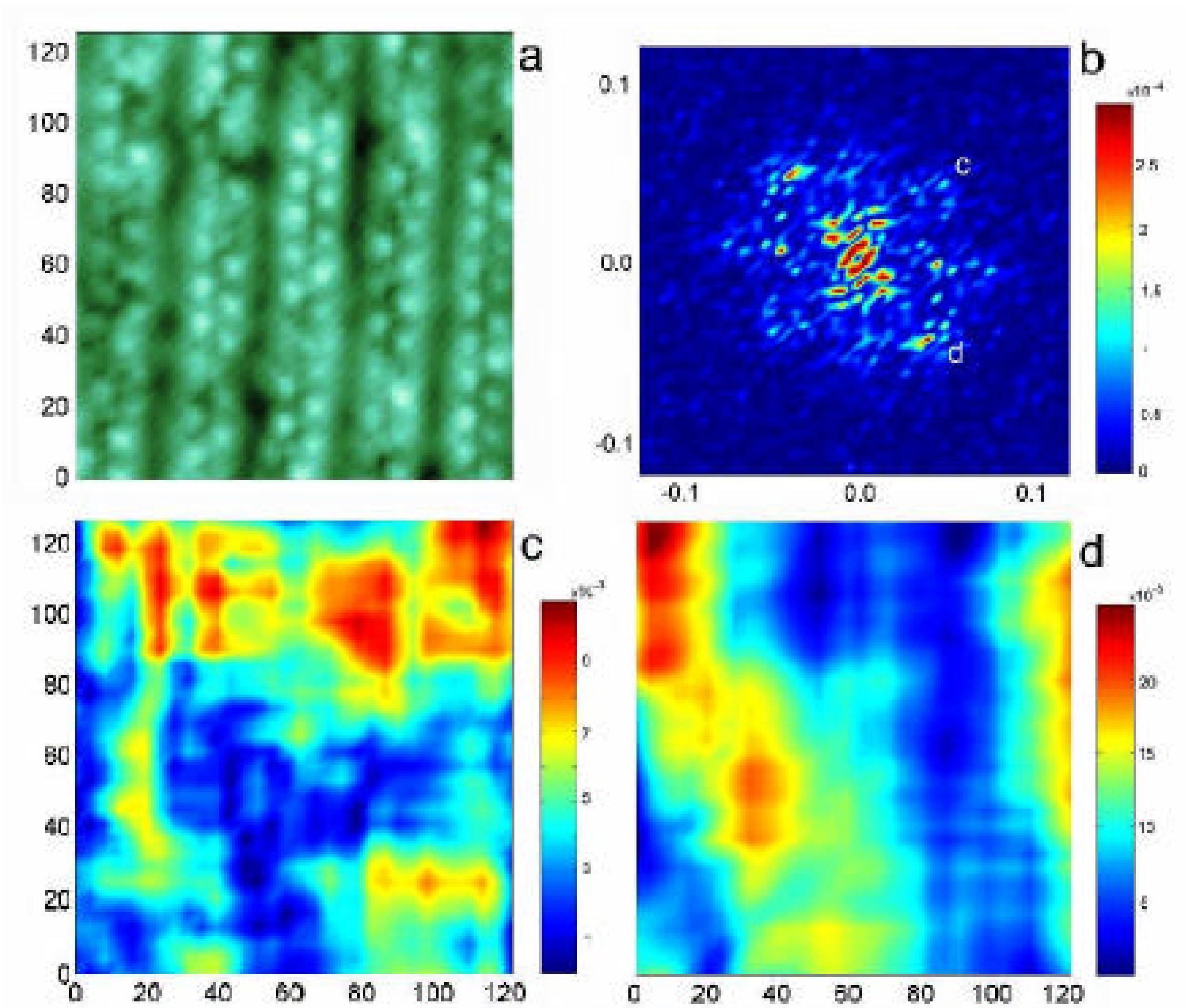

Figure 4